Corresponding Author: Mr Igor Esau, Ph.D.

Corresponding Author's Institution: Nansen Environmental and Remote Sensing Centre

First Author: Igor Esau, Ph.D.

Order of Authors: Igor Esau, Ph.D.; Igor Esau, Ph.D.; Øyvind Byrkjedal, Ph.D.

Abstract: Large-eddy simulation (LES) is a well-established numerical technique, which resolves the most energetic turbulent fluctuations in the planetary boundary layers. Averaging these fluctuations, high-quality profiles of mean values and turbulence statistics can be obtained in experiments with well-defined initial and boundary conditions. Hence, the LES data can be beneficial for assessment and optimization of turbulence closure schemes. A database of 80 LES runs (DATABASE64) for neutral and stably stratified planetary boundary layers (PBLs) is applied in this study to optimize the first-order turbulence closure (FOC). Approximations for the mixing length scale and stability correction functions have been tuned to minimise a relative root-mean square error over the entire DATABASE64. New stability functions have correct asymptotes describing regimes of strong and weak mixing found in theoretical exercises, atmospheric data and LES. The correct asymptotes exclude the need for a critical Richardson number in the FOC formulation. Further, we analysed the FOC quality as functions of the integral PBL stability and the vertical model resolution. We show that the FOC is never perfect because the turbulence in the upper half of the PBL is not generated by the local gradients. Accordingly, the parameterized and LES fluxes decorrelate in the upper PBL. With this imperfection in mind, we show that there is no systematic quality deterioration of the FOC in strongly stable PBLs provided the vertical model resolution is better than 10 levels within the PBL. In agreement with previous studies, we found that the quality improves slowly with the vertical resolution refinement. It is generally wise to not overstretch the mesh in the lowest 500 m of the atmosphere where majority of the observed, simulated and theoretically predicted stably stratified PBLs are located.

Response to Reviewers: Below is our response on specific comments of Reviewer 1.

1. The real PBL is under influence of many unresolved factors, processes and phenomena of which the katabatic flow is just one, perhaps very significant, example. We believe however that flow non-stationarity and flow interaction with unresolved surface heterogeneity are at least equally important. In these circumstances, the considered stationary and homogeneous turbulence closure and its LES validation are seemed to be a rather academic problem. However, the current modeling practice, limited by computer power and understanding of unresolved process in interaction, is essentially connected to such an academic problem. Attempts to introduce elements of heterogeneity into the turbulence closure were not a big success, as to pose it politely, either for meteorological or for LES models.

2. We do not think that Monin-Obukhov scale is so inaccurate. It derived from rather robust energetic considerations. It does not characterize individual eddies but the 40 years of its application has shown that it is indeed imported limited parameter. Derbyshire (1999) for instance used it to derive the maximum possible negative heat flux at the surface, the estimations has been later confirmed with relatively high accuracy by Jimenez et al. using SABLES98 data. So we do believe that to estimates, L and H, are comparable quantities but not estimates of order of magnitude of the relevant scales.

3. We have clarified under revision that "non-local" in this context means "not defined by linear flux-down-gradient relationships". It includes both the mechanisms related on significant turbulent transport of fluxes and irradiation of gravity waves in the stably stratified media. Eq. (13) is not perfect for estimation of the non-locality but it gives useful insight on the FOC performance. In our formulation, "non-local" and "not represented" in the down-gradient FOC have similar meaning.

4. The jet as inertial oscillation is a result of imbalance between constant forcing and initial conditions in our LES. We try to set up them in balance from the start of simulations, so no significant jet could be found in DATABASE64. At the same time, even insignificant jets may produce maxima on the wind profile, which cannot be accounted for in the FOC and so deteriorate correlations.

5. What our Reviewer has attributed as "speculative comments" are in Section "Discussion", which means we do not have enough data or logical constructions to prove those hypothesis. Indeed, we could be wrong in our explanation but it does not contradict our data after all.

6. Nieuwstadt's local theory with linear universal functions versus our universal exponential functions is indeed interesting question. Yes, in the bulk of PBL the linear function seem to be better approximation of the fluxes. Observe however that Taylor expansion of the exponent is a linear function up to the first order approximation. At the top and bottom of the PBL, the linear approximation is physically unacceptable as it leads to discontinuity of the flux profiles at boundaries. Exponents however match the boundaries smoothly.

So it seems to us that exponents have more physical sense then linear approximation even if we cannot construct closed theory for them as Nieuwstadt did.

Running head:
  LES Application to Optimization of First-Order Closures

Article type:
The submission to a special issue of the "*Boundary-Layer Meteorology*" devoted to the NATO advanced research workshop "*Atmospheric Boundary Layers: Modelling and Applications for Environmental Security*"

# Application of a Large Eddy Simulation Database to Optimization of First Order Closures for Neutral and Stably Stratified Boundary Layers

*Igor N. Esau*[1,2,*], *Øyvind Byrkjedal* [2,3]

[1] Nansen Environmental and Remote Sensing Centre, Thormohlensgate 47, 5006, Bergen, Norway
[2] Bjerknes Centre for Climate Research, Bergen, Norway
[3] Geophysical Institute, University of Bergen, Bergen, Norway
[*] corresponding author; e-mail igore@nersc.no

Revisited version R1 from 19 January 2007

# Application of a large eddy simulation database to optimization of first order closures for neutral and stably stratified boundary layers

Igor N. Esau [1,2,*], Øyvind Byrkjedal [2,3]

[1] Nansen Environmental and Remote Sensing Centre, Thormohlensgate 47, 5006, Bergen, Norway

[2] Bjerknes Centre for Climate Research, Bergen, Norway

[3] Geophysical Institute, University of Bergen, Bergen, Norway

[*] Corresponding author; e-mail igore@nersc.no

## Abstract

Large-eddy simulation (LES) is a well-established numerical technique, which resolves the most energetic turbulent fluctuations in the planetary boundary layers. Averaging these fluctuations, high-quality profiles of mean values and turbulence statistics can be obtained in experiments with well-defined initial and boundary conditions. Hence, the LES data can be beneficial for assessment and optimization of turbulence closure schemes. A database of 80 LES runs (DATABASE64) for neutral and stably stratified planetary boundary layers (PBLs) is applied in this study to optimize the first-order turbulence closure (FOC). Approximations for the mixing length scale and stability correction functions have been tuned to minimise a relative root-mean square error over the entire DATABASE64. New stability functions have correct asymptotes describing regimes of strong and weak mixing found in theoretical exercises, atmospheric data and LES. The correct asymptotes exclude the need for a critical Richardson number in the FOC formulation. Further, we analysed the FOC quality as functions of the integral PBL stability and the vertical model resolution. We show that the FOC is never perfect because the turbulence in the upper half of the PBL is not generated by the local gradients. Accordingly, the parameterized and LES fluxes decorrelate in the upper PBL. With this imperfection in mind, we show that there is no systematic quality deterioration of the FOC in strongly stable PBLs provided the vertical model resolution is better than 10 levels within the PBL. In agreement with previous studies, we found that the quality improves slowly with the vertical resolution refinement. It is generally wise to not overstretch the mesh in the lowest 500 m of the atmosphere where majority of the observed, simulated and theoretically predicted stably stratified PBLs are located.

## 1. Introduction

Modelling of a planetary boundary layer (PBL) – the lowest atmospheric layer with significant vertical turbulent diffusion – is a challenge for meteorological simulations. Models need a kind of simplified turbulence-closure schemes to account for complex physical processes and to approximate strongly curved, non-linear mean vertical profiles of meteorological variables in this layer. As any simplified model of physics, a turbulence closure scheme needs validation of assumptions and constraining of involved empirical constants and universal functions. This has been done in a very large number of studies involving data from atmospheric observations (e.g. Businger et al., 1971; Louis, 1979; Businger, 1988; Högström, 1988) and more recently involving numerical data from large-eddy simulations (LES). A comprehensive benchmarking of the closure schemes requires a data set covering the entire parameter space in which models are intended to operate. Although recent observational efforts such as SHEBA (Uttal et al., 2002) and CASES-99 (Poulos et al., 2002) provided relatively accurate data throughout the surface layer and sometimes the entire PBL, any direct intercomparisons are difficult to interpret. The difficulties are lurking in our inability to control the PBL evolution and its governing parameters. Therefore the performance of parameterizations in the benchmarking considerably depends on modellers' skills and understanding of the initial and boundary conditions in the test runs.

In these circumstances, the benchmarking against LES data could be a useful exercise as shown in pioneering works by Deardorff (1972) and Moeng and Wyngaard (1989). The LES data are advantageous as they provide the whole range of turbulence statistics throughout the entire PBL under controllable conditions. Moreover, the LES could be conducted in idealized conditions, which are consistent with the background assumptions behind the closure schemes. It excludes the great deal of irregular discrepancies between the closure and data. The latter sometimes requires fuzzy explanations referring to non-stationary, transitional and meso-scale phenomena recorded in the atmospheric data. It is not surprising then that the LES data look attractive. The pioneering works have been extended in a number of following studies (e.g. Holtslag and Moeng, 1991; Andren and Moeng, 1993; Galmarini et al., 1996; Xu and Taylor, 1997; Nakanishi 2001; Noh et al., 2003). The most recently GEWEX Atmospheric Boundary Layer Study (GABLS) intercomparison study, described in Beare et al. (2006) and Cuxart et al. (2006, hereafter C06) and available on www.gabls.org, involved 11 different LES codes and 20 different turbulence closures.

The GABLS is a case study however. It has little to tell about sensitivity of closures to variations of external parameters. Ayotte et al. (1996) has shown that the closure performance can change significantly for neutral and convective PBLs. Their work was based on intercomparisons between 10 LES runs and corresponding runs of 7 single-column models with different turbulence closures. Up to our knowledge, a similar multi-case study has not been done yet for stably stratified PBLs. This is perhaps due to the considerable computational cost of the stably stratified LES runs. Based on our experience, about two orders of magnitude more computer time is needed to obtain accurate LES data for a typical nocturnal clear-sky PBL as compared to time needed to run LES for a truly neutral PBL. This is due to strong reduction of characteristic turbulence scales both internal and integral followed by the need of adequate mesh refinement and strong reduction of the model time step.

We are the most interested in climate and global scale applications. Although higher-order turbulence-closure schemes gradually become more affordable in the meteorological simulations especially in meso-meteorological ones (e.g. Tjernström et al., 2005), the majority of global-scale models (e.g. five of seven operational models in C06) still rely on simpler, first-order closures (FOC) similar to those in Holt and Raman (1988) review.

To study the FOC performance, a single-column model is usually initiated with more or less relevant vertical profiles of velocity and temperature. Then iterations continue until a quasi-stationary steady state solution is achieved. In this model state, both the mean variables and the turbulent fluxes differ from the data used for comparisons as C06 have clearly shown. It is therefore difficult to trace down reasons for differences and to identify a specific component of the FOC responsible for the deviations. Here, we will not study an equilibrium solution. Instead we will test different components of the FOC against data from a LES database of 80 neutral and stably stratified runs, hereafter DATABASE64. It is reasonable to expect that, given the stationary, steady state mean profiles for the wind speed and temperature, an ideal parameterization would recover fluxes and tendencies similar to those from DATABASE64. Any discrepancy would be an indication of an internal inconsistency in the parameterization.

The performance of the FOC is also expected to change with vertical mesh resolution. In C06, the performance is studied at resolution of 6.25 m. Realistic resolutions are much coarser. There are typically 2 to 4 model levels within the stably stratified PBLs in actual simulations. Such a coarse resolution could significantly alter the FOC performance as the closure relies on vertical gradients computed on a finite mesh. For instance, Lane et al. (2000) showed considerable convergence of the single-column model results on consequently refined meshes. More recently, Roeckner et al. (2006) reported significant improvement in representation of the near-surface high-latitude climate with resolution refinement in the ECHAM5 model. In contrast, C06 quoted results fairly insensitive to the vertical resolution for the mesh spacing less than 50 m. Ayotte et al. (1996) results suggest a way to resolve this contradiction. They showed a considerable improvement of accuracy for the resolution up to 10 levels in the PBL. It gives the spacing of about 50 m for a typical conventionally neutral PBL but the spacing should be much finer to resolve a long-lived stably stratified PBL (Zilitinkevich and Esau, 2003).

The structure of this paper is as follows. Section 2 describes the large-eddy simulation code and the DATABASE64. Section 3 repeats a description of a general first-order closure and provides details of the optimized FOC fitted to DATABASE64. Section 4 gives the FOC quality assessment over the range of stably stratified PBL and resolutions. Section 5 discuss possible effects of non-local fluxes on the FOC performance. Section 6 outlines conclusions.

**2. Large-eddy simulation data**

Turbulence-resolving simulations have been conducted with the Nansen-centre large-eddy simulation code LESNIC. The code solves three-dimensional momentum, temperature and continuity equations for incompressible Boussinesq fluid. It employs a fully conservative 2nd order central difference scheme for the skew-symmetric advection term; the 4th order Runge-Kutta scheme for time stepping; and a direct fractional-step pressure correction scheme for the continuity preservation. The computational mesh is the staggered C-type mesh, which demands only fluxes as boundary conditions. The LESNIC employs a dynamic mixed closure, which recalculates the Smagorinsky constant in the eddy diffusivity part of the closure at every time step. Detailed description of the LESNIC was published in Esau (2004), intercomparisons – in Beare et al. (2006) and Fedorovich et al. (2004).

The LESNIC has been used to compute a set of cases, referred to as DATABASE64 (Esau and Zilitinkevich, 2006). In DATABASE64, 80 LES runs are suitable for the present study. All runs have been computed at the equidistant mesh with $64^3$ grid nodes. The aspect ratio between the vertical and horizontal grid spacing varies from 1:1 to 1:4 with the majority of data computed at the 1:2 mesh. The physical resolution varies from run to run keeping about 25 to 45 vertical levels within the fully developed PBL.

The turbulent boundary layer comprised only 1/2 to 2/3 of the depth of the computational domain. This arrangement assures that the largest eddies, which occupy the entire PBL, were not affected by the limited horizontal size of the LESNIC domain. The lateral boundary conditions were periodic in all runs. At the surface, the turbulent flux of potential temperature was prescribed and therefore it is considered in this study as an external parameter; whereas the turbulent flux of momentum was computed instantly and pointwise using the log-law. With these boundary conditions, turbulent surface flux parameterizations cannot be evaluated accurately as the surface temperature has not been prescribed in the LESNIC but could be retrieved from the data only with some type of parameterizations.

The vertical profiles of the mean variables and the turbulence statistics have been computed from instant resolved-scale fluctuations. For instance, a vertical flux of the potential temperature, $\tau_\theta$, is computed as

$$\tau_\theta(z) = \overline{<w'\theta'>} + \overline{\tau_{SGS}} = \overline{<(w(x,y,z,t)-<w(x,y,z,t)>)(\theta(x,y,z,t)-<\theta(x,y,z,t)>)>} + \overline{\tau_{SGS}},$$

where $w(x,y,z,t), \theta(x,y,z,t)$ are instant values of the vertical component of velocity and the potential temperature at every grid node. The angular brackets denote horizontal averaging over the computational domain. The overbar denotes time averaging. All runs in DATABASE64 have been calculated for 16 hours of model time. Only the last hour was used for the time averaging. A sub-grid scale flux, $\tau_{SGS}$, is provided from the dynamic mixed closure employed in the LESNIC. Comparing DATABASE64 with the LESNIC runs at $128^3$ mesh and at $256^3$ mesh (for one truly neutral case), we estimated that about 80% of the fluctuation kinetic energy and 65% of the vertical velocity variations at the first computational level were resolved. Thus, $\tau_{SGS}$ is about 20% to 40% of $\tau_\theta$ at the first level and reduces with height.

## 3. Optimization of the first-order turbulence closure

*3.1. Description*

A prognostic tendency due to the turbulent flux divergence in a meteorological model (e.g. Holt and Raman, 1988; C06) can be expressed as

$$\frac{\partial \psi}{\partial t} = -\nabla_z \tau_\psi, \qquad (1)$$

where $\psi$ is one of prognostic variables, and $\tau_\psi$ is its vertical turbulent flux. To avoid model specific details, we consider the closure in Eq. (1) as applied to a horizontally homogeneous boundary layer of incompressible, barotropic fluid flow with a constant turbulent temperature flux at the surface. Such conditions describe realistic shallow stably stratified PBL. The vertical flux $\tau_\psi$ is parameterized following a flux-gradient assumption (Businger et al., 1971; Deardorff, 1972) through an eddy diffusivity, $K_\psi$, and the vertical gradient of mean variables as

$$\tau_\psi = K_\psi \nabla_z \psi \qquad (2)$$

Eddy diffusivity parameterizations generally follow Louis (1979). There is however a rich variety of individual variations in different models. A general form reads

$$K_\psi = l^2 \left| \nabla_z \vec{U} \right| f_\psi \qquad (3a)$$

Another, non-local formulation (e.g. Troen and Mahrt, 1986) based on predefined shape of $K_\psi$-profile is also popular. For cubic profiles, it reads

$$K_\psi = \kappa \tau_0^{1/2} \phi_\psi^{-1} z(1 - z/H)^2, \qquad (3b)$$

where $\phi_\psi = 1 + C_\psi z/L$; $f_\psi$ is a stability correction function; $l$ is a mixing length scale; $C_\psi$ is an empirical constant; $z$ is the height above the surface; and $H$ is the boundary layer depth. Both $f_\psi$ and $l$ could be universal functions of the gradient Richardson number, $Ri = g/\theta_0 \nabla_z \theta (\nabla_z \vec{U})^{-2}$.

The mean variables are: $\vec{U}$ – the horizontal wind velocity; and $\theta$ is the potential temperature. $L = -\tau_0^{3/2}/g\theta_0^{-1}\tau_{\theta 0}$ is the Monin-Obukhov length scale. The index 0 denotes surface values of the momentum, $\tau(z) = \sqrt{<u'w'>^2 + <v'w'>^2}$, and temperature, $\tau_\theta(z) = <\theta'w'>$, fluxes. The constants $\kappa, g, \theta_0$ are the von Karman constant, 0.41; the gravity acceleration, 9.8 m s$^{-2}$, and the reference potential temperature, 300 K. Theory and observational data suggest that the stability functions $f_m, f_\theta$ for momentum and temperature should differ asymptotically as the ratio of the corresponding turbulent fluxes, known as the Prandtl number Pr, changes with $Ri$. However, there are no obvious reasons for $l$ to be different for momentum and the temperature as is often assumed in C06 parameterizations.

*3.2. Fitting of empirical functions*

There is a great variety of parameterizations for $f_\psi$ and $l$ as they have been thought to be the best objects for modifications. Surveys of the proposed functions can be found in Holt and Raman (1988), Derbyshire (1999), Mahrt and Vickers (2003) and C06. It is interesting to determine these functions and to assess their universality using data from DATABASE64. An original concept behind distinction between $f_\psi$ and $l$ was to use $l = l(z)$ to account for the eddy-damping effect of an impenetrable wall (Nikuradze, 1933; Blackader, 1962) and to use $f_\psi = f_\psi(Ri)$ to account for the effect of the static stability (Djolov, 1973; Louis, 1979). We will refer here only to the original concept excluding from the discussion the existing multitude of $l$ modifications, which involve stability corrections (e.g. Mahrt and Vickers, 2003). Figure 1 shows the mixing length scale obtained in truly neutral runs as

$$l = \tau_\psi^{1/2} \left( \nabla_z \vec{U} \middle| \nabla_z \vec{U} \right)^{-1/2}. \tag{4}$$

The mixing length scale is indeed universal function in the lower half of the PBL. It could be approximated with a simple relationship following Blackader (1962) as

$$l = (1/\kappa z + 1/l_0)^{-1}, \tag{5}$$

where $\kappa = 0.41$ is the von Karman constant. It is still used in many models. The best asymptotic fit of $l_0/H$ to the DATABASE64 (Figure 2) is $l_0/H = 0.3$. This is surprisingly close to the empirical best fit in some meso-scale models (e.g. Ballard et al., 1991). The $l_0/H$ ratio gives 100 m $< l_0 <$ 200 m for the near-neutral PBLs, which includes the frequently quoted value $l_0 = 150$ m (Viterbo et al., 1999). The mixing length scale exhibits irregular fluctuations in the middle of the PBL. These fluctuations are thought to be due to the inflection point on the velocity profile where the mean velocity gradient is close to zero. It could be feature related to roll structures (Brown, 1972) or the low-level jet development (Thorpe and Guymer, 1977). Thus, the large values of $l$ could be due to inconsistency between the local flux-profile assumption and more complicated, non-local nature of the PBL turbulence. As we will see much of the FOC inaccuracy could be regarded to this inconsistency. To address literature on $l_0$ (e.g. Weng and Taylor, 2003), we compute the following equation

$$l_0 = C_l q_n = C_l \frac{\int_0^\infty z q \, dz}{\int_0^\infty q \, dz}, \quad q = \sqrt{2E}, \tag{6}$$

where $E$ is the turbulent kinetic energy and $C_l$ is an empirical constant. Weng and Taylor cited the range $0.1 < C_l < 0.25$ from literature but argued that $C_l$ should be as small as 0.055. Using rather good proportionality $q_n = 0.625 H$ (see Figure 3), we determine much larger optimal value $C_l = (1/0.625) l_0 / H = 0.48$ from DATABASE64. Such $C_l$ would require retuning of all other constant in the turbulence closures since it is an order of magnitude larger than the optimal value found by Weng and Taylor.

Using the fitted mixing length scale from Eq. (5) in combination with Eq (2) and Eq(3a), one can compute the stability correction functions from DATABASE64 as

$$f_\psi = \frac{\tau_\psi}{\left|\nabla_z \vec{U}\right|\nabla_z \psi (1/\kappa z + 1/l_0)^{-2}}. \tag{7}$$

These functions for momentum and temperature as well as their ratio, known as the Prandl number, are shown in Figure 4. The presented LES data behave similarly to SABLES98 data shown in Yague et al. (2006). It is clearly seen that popular approximations in C06 are in rather poor agreement both with DATABASE64, a high-resolution LESNIC run for the GABLS test case (Beare et al., 2006) and with theoretical considerations presented by Zilitinkevich and Esau in this issue. An empirical fit based on the bin-averaged DATABASE64 data reads

$$f_m = (1 + a_m Ri)^{-2} + b_m Ri^{1/2}, \tag{8a}$$

$$f_\theta = (1 + a_\theta Ri)^{-3} + b_\theta, \tag{8b}$$

where $a_m = 21$, $b_m = 0.005$, $a_\theta = 10$, $b_\theta = 0.0012$. Interesting that similar functions with exponents $-2$ and $-3$ correspondingly have appeared in Derbyshire (1999) study. These functions demonstrated physically consistent behaviour in equilibrium runs of a single-column model for the stably stratified PBL. No decoupling between the surface and the atmosphere has been observed even at the strongest stabilities. These functions in Eq. (8) almost perfectly approximate the Prandtl number with an inflection point, which is a kin to the critical Richardson number in the Zilitinkevich's total turbulent energy theory (see this issue), at $Ri_{cr} = 0.15$. The data reveal two regimes of strong continues turbulence at $Ri < Ri_{cr}$ and weak, probably intermittent, turbulence at $Ri > 1$ with transitional zone $0.1 < Ri < 1$. The higher resolution GABLS run (darker grey dots in Figure 4) provides data within the scatter range of the DATABASE64 but with some systematic deviations from the bin-averaged values. It may indicate insufficient resolution of the DATABASE64 or data dependence on evolution of the capping inversion (the layer with the largest $Ri$). It is worth to emphasise that the GABLS run was computed for 9 model hours while the DATABASE64 for 16 model hours. Only the last model hour was used in this analysis. Hence, the capping inversion in the GABLS run could be underdeveloped.

Figure 4 discloses substantial inconsistency between DATABASE64 and the formulations for the stability correction functions in the meteorological models. Small revisions of the Louis (1979) formulations do not improve the approximations, which have wrong asymptotes at large $Ri$ and therefore the incorrect Prandtl number.

Figure 5 helps to determine value of the constants $C_\psi$ in Eq. (3b) for the non-dimensional gradients $\phi_\psi$. Commonly cited values are between 4 and 7 (e.g. Högström, 1988). DATABASE64 gives a rather certain estimation for momentum where $C_u$ = 5.9 (= 2.5 in terms of Zilitinkevich and Esau (2005) formulation). As expected from the Prandtl number behaviour, the constant $C_\theta$ = 7.3, matched to our LES data, does not fit the proposed linear dependence. According to Zilitinkevich and Esau (2005) both constant should appear in Eq. (3b) where a generalized turbulence length scale is used instead of the surface Monin-Obukhov length scale. It is clearly seen that the linear approximation in Eq. (3b) for the temperature non-dimensional gradient is not suitable, as it would lead to a limitation Pr < $C_\theta/C_u$ < 2 on the Prandtl number. Advanced theory (Zilitinkevich, personal communication) predicts quadratic dependence on the non-dimensional height.

## 4. Quality assessment

*4.1. Quality dependence on turbulence structure*

Figure 6 shows vertical profiles of correlations between the FOC fluxes and the directly calculated fluxes from DATABASE64. Correlations at a few lowest layers are affected by the prescribed surface fluxes. Above this layer, there is generally good agreement between DATABASE64 and the FOC in the lower half of the PBL. The mean correlations are as large as 0.9 for all cases. The correlations deteriorate in the upper PBL with minimum of about 0.2 at the PBL top.

Detailed analysis reveals that the temperature flux decorrelates only for the conventionally neutral and the long-lived PBLs with significant temperature inversions at the PBL top. In the inversion layer, even small eddies are able to produce temperature fluctuations in presence of large mean temperature gradients. This flux is not accounted for in the FOC. C06 showed that the majority of turbulence closures failed to develop even a relatively weak inversion in the GABLS case after 9 hours of simulations. Similar conclusion has been drawn from intercomparisons between MODIS satellite data and NCEP reanalysis data (Liu and Key, 2003), which are the most consistent data large-scale model can produce. The correlations remain considerably higher in the upper nocturnal PBL where the entrainment flux is insignificant. The average correlation is about 0.8 for the temperature flux but 0.4 for the momentum flux. These results are in general agreement with Galmarini et al. (1998) conclusions. They demonstrated nearly perfect reproduction of a weakly stratified nocturnal PBL by a second-order turbulence closure in comparisons with two LES.

*4.2. Quality dependence on stability*

Even the optimized FOC is never perfect. A relative root-mean square error can be calculated as

$$\varepsilon_\psi^{RMS} = \left( \sum_{n=1}^{N_z} \left( \tau_\psi^{FOC}(z_n) - \tau_\psi^{LES}(z_n) \right)^2 \right)^{1/2} \cdot \left( \sum_{n=1}^{N_z} \left( \tau_\psi^{LES}(z_n) \right)^2 \right)^{-1/2}, \qquad (9)$$

where superscripts "FOC" and "LES" denote the fluxes from the FOC initiated with the mean profiles from DATABASE64 and the fluxes computed directly in the LES; $N_z$ is the number of LES levels within the PBL and $z_n$ is the height of the n$^{th}$ level. After 15 hours of integration, the runs in DATABASE64 provide quasi-stationary, steady state mean variables and fluxes. Hence it is reasonable to expect consistency between the FOC and the LES fluxes computed from the equilibrium temperature and wind profiles, i. e. $\varepsilon_\psi^{RMS} \to 0$. Figure 7 shows this is not the case. The error does not depend systematically on the PBL thickness, which is an integral measure of the PBL stability (Zilitinkevich and Esau, 2003). The error is larger only for the cases, which are absolutely dominated by the entrainment fluxes from the capping inversions. These cases are particularly difficult for the FOC to reproduce. The minimum asymptotic error for DATABASE64 is $\varepsilon^{LES} > 0.05$ both for momentum and temperature fluxes. Existence of a rather large minimum asymptotic error has also been found by Ayotte et al. (1996).

*4.3. Quality dependence on vertical resolution*

Such a fine vertical resolution as in DATABASE64 is unattainable in the meteorological models. The practically sounded resolution varies from 4 to 7 levels for relatively deep, weakly stratified PBL (Tjernström et al., 2005). The resolution deteriorates to a single level for polar long-lived and strongly stratified PBL (Cassano, 2001). Ayotte et al. (1996) found that $\varepsilon^{LES}$ is reached in neutral and convective PBL at resolutions $N_z > 10$. Our Figure 8 partially supports their conclusion. The error slowly decays with resolution refinement. The largest quality improvement is however seen in the interval $10 < N_z < 15$. To obtain Figure 8, we have computed $\varepsilon_\psi^{RMS}$ for a number of regular meshes with inter-level spacing 5 m, 10 m, 15 m, 20 m, 30 m, 50m and 70 m for every run from DATABASE64. Then the average $\varepsilon_\psi^{RMS}$ for every $N_z$ was computed with subtraction of own $\varepsilon^{LES}$ for every LES run.

Another important aspect of the vertical resolution is a numerical approximation of strongly curved vertical profiles on coarse meshes using finite-difference numerical schemes. In majority of models, calculations of the FOC require calculations of the vertical gradients with the 2$^{nd}$ order central differences as e.g. for the vertical flux divergence below

$$\nabla_z^\delta \tau_\psi = \delta^{-1}\left(\tau_\psi(z_n) - \tau_\psi(z_{n-1})\right), \qquad (10)$$

where $\delta = z_n - z_{n-1}$ is the distance between n$^{th}$ and n$^{th}$-1 model levels at which the variable $\psi$ is found.

We can quantify distortions introduced by the finite-difference operator $\nabla_z^\delta$ in the non-linear universal non-dimensional profiles of the turbulent fluxes (Zilitinkevich and Esau, 2005). Proposed analytical approximations of these profiles read

$$\tau(z)/\tau_0 = \exp(-8/3(z/H)^2), \qquad (11a)$$

$$\tau_\theta(z)/\tau_{\theta 0} = \exp(-2(z/H)^2), \qquad (11b)$$

where $\tau_0, \tau_{\theta 0}$ are surface values of the momentum and temperature fluxes correspondingly. DATABASE64 data and the analytical functions are shown in Figure 9. The vertical profiles of the relative distortions in velocity and temperature tendencies can be computed as

$$\varepsilon_\psi(z) = \frac{\left(\frac{\partial \psi}{\partial t}\Big|_\delta - \frac{\partial \psi}{\partial t}\right)}{\frac{\partial \psi}{\partial t}} = \frac{(\nabla_z^\delta \tau_\psi - \nabla_z \tau_\psi)}{\nabla_z \tau_\psi}, \qquad (12)$$

Figure 10 shows the profiles of $\varepsilon_\psi(z)$ for different regular meshes with $N_z$ stated on the plot. One can notice that the operator $\nabla_z^\delta$ introduces a negative relative error in the model tendencies within the PBL and a positive error immediately above the PBL. The same must be true for the Richardson number and for the fluxes. To keep the differentiation errors reasonably within 10% of accuracy, the model should have $N_z > 6$.

**5. Discussion**

The presented technical data point out on an important role of non-local turbulent mixing within the stably stratified PBL. The definition "non-local" is used here in a broad sense indicating inconsistency between the FOC assumption of a linear flux-gradient correlation and the observed de-correlations disregarding the physical mechanisms behind these inconsistencies. It seemingly contradicts to the commonly accepted hypothesis that the turbulent eddies in the stably stratified PBL are small and therefore their generation-dissipation is governed by the local gradients. In fact this local mixing hypothesis is supported only by data from the lower nocturnal PBL (Nieuwstadt, 1984) where the capping inversion is relatively weak. In this case the turbulence scale is limited by the local Monin-Obukhov length scale $\Lambda = \tau^{3/2}/\kappa g \theta_0^{-1} \tau_\theta$ but the PBL thickness is larger than the mean eddy size, i.e. $\Lambda < H$. Zilitinkevich (2002) suggested that in many cases the stability of the free atmosphere imposes stronger limitations on the turbulence scale than $\Lambda$. Zilitinkevich and Esau (2002; 2003) and Zilitinkevich et al. (2006) demonstrated this with the LES data. In the atmosphere, the cases with $\Lambda > H$ are frequently observed in mid- and high-latitudes during wintertime where the strong negative radiation imbalance and the large-scale air subsidence in anti-cyclones play together to create and strengthen temperature inversions (Overland and Guest, 1991). A typical long-lived stably stratified PBL, as it has been observed at the SHEBA site in Arctic is between 100 m to 200 m thick and capped by a relatively strong potential temperature inversion with $\nabla_z \theta \sim 6$ K km$^{-1}$. It is rather unusual to observe (and very difficult to simulate) situations with $\Lambda < H$ under conditions with any significant wind speed.

Following Deardorff (1972) and Holtslag and Moeng (1991), one can estimate relative importance of the non-local, counter-gradient temperature flux in the entire stably stratified PBL as

$$R_\theta = \sum_z \frac{\tau_\theta^{local}}{\tau_\theta^{total}} = \sum_z \left(1 - \frac{\gamma_\theta}{\nabla_z \theta}\right), \qquad (13)$$

where $\gamma_\theta = g\theta_0^{-1}\overline{<\theta'\theta'>}/\overline{<w'w'>}$; and $\tau^{total}$ and $\tau^{local}$ are total and local, i.e. not down-gradient, fluxes. This measure is not perfect. It just attempts to fit the complex turbulence physics to one-dimensional flux-profile theory. Nevertheless, it clearly shows (see Figure 11) the importance of the non-local flux. The non-local flux dramatically increases for $\Lambda < H$ both in the lower and the upper parts of the PBL. Moreover, the non-local flux in the upper PBL is always more important for the total turbulence mixing than in the lower PBL. Hence DATABASE64 data only partially support Holtslag and Boville (1993) assumption that $\gamma_\theta$ is small and could be neglected in stable conditions.

As we see it, the FOC fails due to two different physical reasons. Consider the flux in the FOC as multiplication of the three-dimensional eddy diffusivity characterizing the turbulent mixing and the three-dimensional mean gradient of a meteorological quantity. In one asymptotic regime of the nocturnal PBL, the gradient is created through the turbulent mixing, which compensates the surface cooler temperature and the surface friction. Thus, the gradient is dependent on the mixing. The mixing is however suppressed as the strongest gradients are observed at the surface. Pressure variations and eddy interactions with the mean flow result in development of eddy activity in the layers with small or no gradients. In this layer at the PBL top, one can observe mixing, as DATABASE64 shows, which is considerably stronger than it could be if it was simply imported from the lower layers. In fact the transport terms in the TKE budget are fairly small (~ $10^{-5}$ m$^2$ s$^{-3}$ and ~ $10^{-6}$ K$^2$ s$^{-1}$) in the upper nocturnal TKE. In another asymptotic regime of conventionally neutral PBL, the gradient is imposed through free atmosphere stability. Thus, the mixing, imported in the PBL top from lower weakly stratified layers, is stronger than the locally generated mixing. The TKE transport terms increase by factor 10 or more in comparison with the nocturnal PBL.

As we have seen the PBL top and the capping inversion are the most difficult layer to parameterize in frameworks of the first order closure. The inversion layer is important as in the high-latitudes the fluxes, especially the temperature and moisture fluxes, formed in that layer, counteract radiative surface cooling, low-level clouds and fog formation. The FOC intrinsically provides too small fluxes, which if they are not corrected, would allow the surface temperature drop down to much lower values than it has been observed. In part, the flux enhancement, disclosed in C06, is to prevent such a PBL decoupling from the surface (Derbyshire, 1999). Figure 12 provides intercomparisons between the LES and the FOC fluxes for several high-resolution runs. Generally, the FOC overestimates fluxes in the lower PBL and underestimates them in the upper PBL. Taking into account that single-column models based on the FOC in C06 predicted deeper PBL than the LES did, it could be argued that in the simulation, the entire observed PBL would be situated within the layer of overestimated fluxes. Small FOC fluxes in $z/H < 0.05$ are numerical artefact since no surface flux parameterization has been employed here.

## 6. Conclusions

In this study, we have applied the large-eddy simulation database DATABASE64 of medium-resolution (25 to 45 levels within the PBL) runs to optimize the simplest but still popular first-order closure and to assess its quality over a range of governing parameters in the stably stratified PBLs. The analysis has been also supported by three higher resolution LES runs, one of which was the GABLS run (Beare et al., 2006). Our LES numerically resolve the most energetic fluctuations of velocity and potential temperature. It allows the use of the high-quality quasi-stationary, steady state mean profiles as input for the flux calculations in the FOC. The FOC fluxes were compared with directly computed turbulent fluxes from DATABASE64.

To benefit from this approach, we optimized the elements of the FOC, such as the mixing length scale and the stability correction functions, to achieve the best agreement with 80 LES experiments for neutral and stably stratified PBLs in DATABASE64. We have demonstrated general inconsistency of the traditionally used constants and universal functions with the LES data. New stability functions have been proposed. They correct asymptotes in the regimes of strong and week mixing in accordance with DATABASE64 and the recent development in turbulence theory and observations during SABLES98 campaign. The correction excludes the need for a critical Richardson number.

We have analysed the FOC quality as a function of the vertical model resolution and the PBL stability. The FOC is never perfect. The reason is that in the upper PBL the mixing or the temperature gradients are essentially non-local. Accordingly, the parameterized and LES fluxes decorrelate in the upper PBL. As in previous studies, we have shown that the quality improves slowly with the vertical resolution refinement. It is generally enough to have at least 10 levels within the PBL to achieve the FOC accuracy close to the asymptotically possible. As even 10 levels could be too costly for the meteorological models, it is wise to not overstretch the mesh in the lowest 500 m of the atmosphere where majority of the observed, simulated and theoretically predicted stably stratified PBLs are located.

Although there could be other reasons as well, we speculated that the common flux enhancement needed to run the models with the FOC is originated by internal model inconsistencies: the coarse vertical mesh, which smoothes vertical gradients; the overmixing in the lower PBL, which biases and smoothes the mean profiles; and undermixing in the upper PBL, which reduces the warm air entrainment. Thus, to satisfy the surface energy budget in the meteorological models, where the turbulent flux is just one of the components, the FOC should provide larger fluxes than it has been defined from the flux-gradient assumption.

**Acknowledgements**

This work has been supported by the Norwegian project POCAHONTAS 178345/S30, joint Norwegian-USA project PAACSIZ 178908/S30. Cooperation in frameworks of the NORDPLUS Neighbour 2005-2007 network FI-51 was essential for this study. The authors thank Prof. N.-G. Kvamstø (Geophysical Institute of Bergen University) and Prof. S. S. Zilitinkevich (Helsinki University) for supporting discussions.

**Figure Captions**

**Figure 1**. Profiles of the normalized mixing length scale, $l/H$, where $H$ is the PBL thickness, computed for the truly neutral LES runs from DATABASE64 (light dots) and for higher resolution $128^3$ and $256^3$ LES runs (dark dots). The solid curve represents the Blackadar (1962) approximation after Eq. (5) and $l_0/H = 0.3$. The dashed line represents Nikuradze (1933) formulation given by $l = -0.06z^4 + 0.24z^3 - 0.2z^2 + \kappa z$.

**Figure 2.** Dependence of the mean relative error given in Eq. (9) from the normalized asymptotic length scale . Dark dots and solid curve show the errors in the momentum flux and the light dots and dotted curve – in the temperature flux.

**Figure 3.** Dependence of the integral normalized TKE on the PBL thickness: light dots – DATABASE64 data; solid line – the best linear fit to the data.

**Figure 4.** Stability correction functions and their ratio (the Prandtl number) as functions of the gradient Richardson number: $f_m(Ri)$ – for momentum (a); $f_h(Ri)$ – for temperature (b); and $\Pr = f_m(Ri)/f_h(Ri)$ – the Prandtl number (c). Symbols are: light dots – DATABASE64 data; dark dots – LES simulations for the GABLS case (Beare et al., 2006); black dots with error bars – bin-averaged values of equally weighted DATABASE64 data, the horizontal bars denote the bin width, the vertical bars denote one standard deviation within each bin for the data found above and below of the bin-average value. Curves are: solid curve – the best fit to the bin-averaged DATABASE64 data; dashed curve – the approximation used in the ECMWF, the ARPEGE (MeteoFrance) and some other models (Louis, 1982; C06); dash-dotted curve – an improved MeteoFrance approximation discussed in C06; dotted curve – the approximation used in the MetOffice model.

**Figure 5.** Constants in flux-profile relationships after Eq. (3b): $C_u$ – for momentum (a) and $C_\theta$ – for the temperature (b) non-dimensional gradients as functions of the height normalized by the surface Monin-Obukhov length scale. Light dots denote DATABASE64 data for the lower 1/3 of the PBL with exception of the surface layer (1-2 computational levels in the LES). The bold lines are the mean values of the constants $C_u = 5.9$ and $C_\theta = 7.3$. It is clearly seen that the linear approximation for the non-dimensional temperature gradient is not suitable.

**Figure 6.** Vertical profiles of the correlations between the turbulent fluxes in the optimized FOC and the directly computed LES fluxes from DATABASE64. Solid curve is for the momentum and the dotted curve – for the temperature flux correlations: (a) – averaged over all cases; (b) – averaged over conventionally neutral cases; (c) – averaged over nocturnal cases. High correlations near the surface, $z/H < 0.05$, are artificial product of the numerical analysis.

**Figure 7.** Relative root-mean square error over the boundary layer after Eq. (9) for the momentum (a) and temperature (b) fluxes computed for the optimized FOC from DATABASE64 wind and temperature profiles.

**Figure 8.** The FOC quality dependence on model resolution within the PBL. The quality computed after Eq. (9) with subtraction of the minimum error for every LES run. Dark bars are for the momentum flux and light bars – for the temperature flux computed by the optimized FOC from the DATABASE64 wind and temperature profiles.

**Figure 9.** Non-dimensional turbulent fluxes for momentum (a) and temperature (b) taken from DATABASE64 (light dots) and their analytical approximations (bold curves) with Zilitinkevich and Esau (2005) exponential universal functions after Eq. (11).

**Figure 10.** Profiles of relative errors in momentum (a) and temperature tendencies (b). The errors are computed by Eq. (12) for different equidistant meshes with $N_z = H/\delta$ indicated at the bottom of the corresponding curve.

**Figure 11.** Ratio between the total and the down-gradient fluxes computed from DATABASE64 after Eq. (13). The ratio is plotted versus non-dimensional ratio between the surface Monin-Obukhov length scale and the PBL thickness. Symbols are: light dots – the averaged values for the lower half of the PBL; dark dots – the averaged values for the upper half of the PBL.

**Figure 12.** Vertical profiles of the LES (solid curves) and the FOC (dashed curves) fluxes for momentum (a, c, e) and temperature (b, d, f) for three distinct cases: (a, b) – the nocturnal PBL with zonal geostrophic wind, $U_g = 5$ m s$^{-1}$, the surface temperature flux $\tau_{\theta 0} = 0.001$ K m s$^{-1}$, at latitude 45 degrees North and surface roughness 0.1 m; (c, d) – the long-lived stably stratified PBL in the GABLS case, with zonal geostrophic wind, $U_g = 8$ m s$^{-1}$, the variable surface temperature flux with the mean value of $\tau_{\theta 0} \sim 0.07$ K m s$^{-1}$, at latitude 73 degrees North and surface roughness 0.1 m; (e, f) – the conventionally neutral PBL with zonal geostrophic wind, $U_g = 5$ m s$^{-1}$, the surface temperature flux $\tau_{\theta 0} = 0$ K m s$^{-1}$, at latitude 45 degrees North and surface roughness 0.1 m. The FOC fluxes in $z/H < 0.05$ are numerical artefact as no surface flux parameterization has been applied.

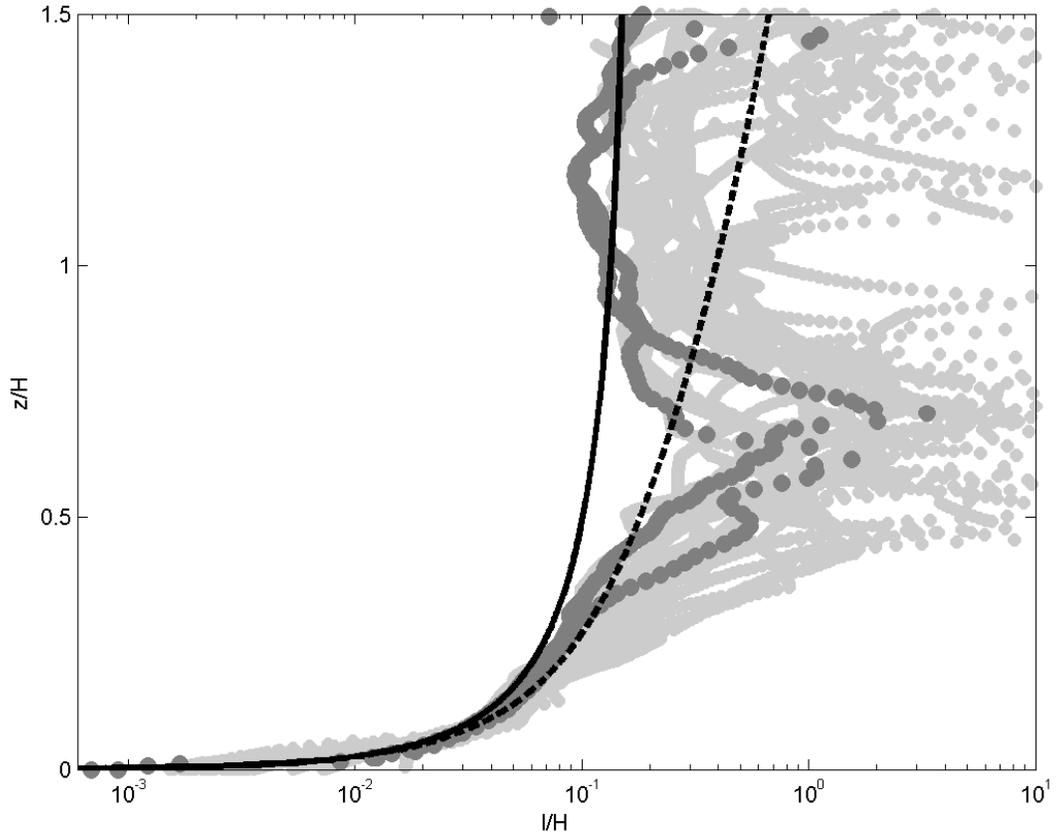

**Figure 1**. Profiles of the normalized mixing length scale, $l/H$, where $H$ is the PBL thickness, computed for the truly neutral LES runs from DATABASE64 (light dots) and for higher resolution $128^3$ and $256^3$ LES runs (dark dots). The solid curve represents the Blackadar (1962) approximation after Eq. (5) and $l_0/H = 0.3$. The dashed line represents Nikuradze (1933) formulation given by $l = -0.06z^4 + 0.24z^3 - 0.2z^2 + \kappa z$.

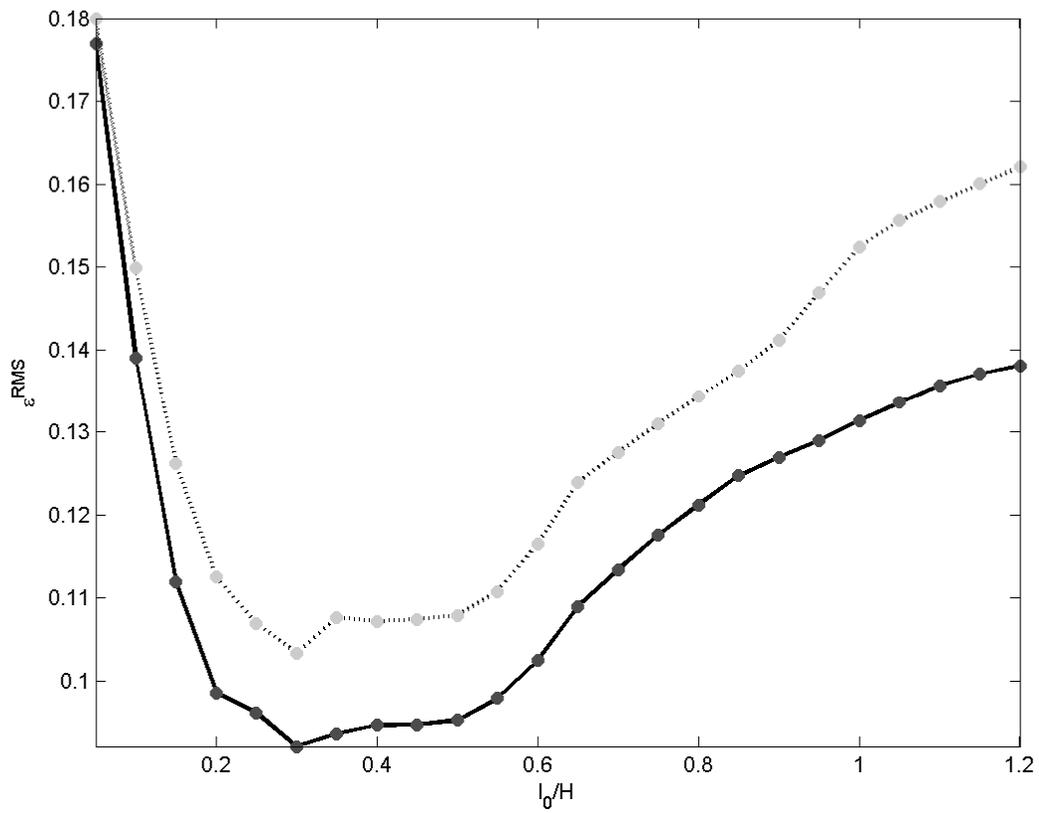

**Figure 2.** Dependence of the mean relative error given in Eq. (9) from the normalized asymptotic length scale $l_0/H$. Dark dots and solid curve show the errors in the momentum flux and the light dots and dotted curve – in the temperature flux.

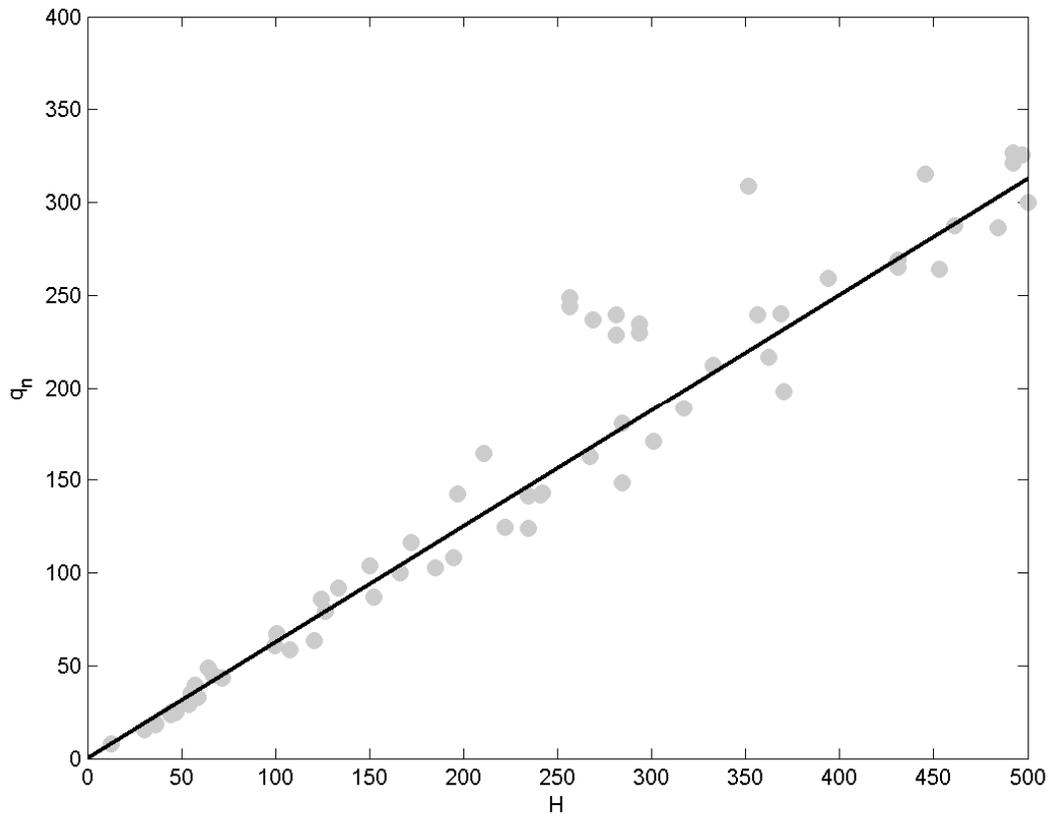

**Figure 3.** Dependence of the integral normalized TKE on the PBL thickness: light dots – DATABASE64 data; solid line – the best linear fit to the data.

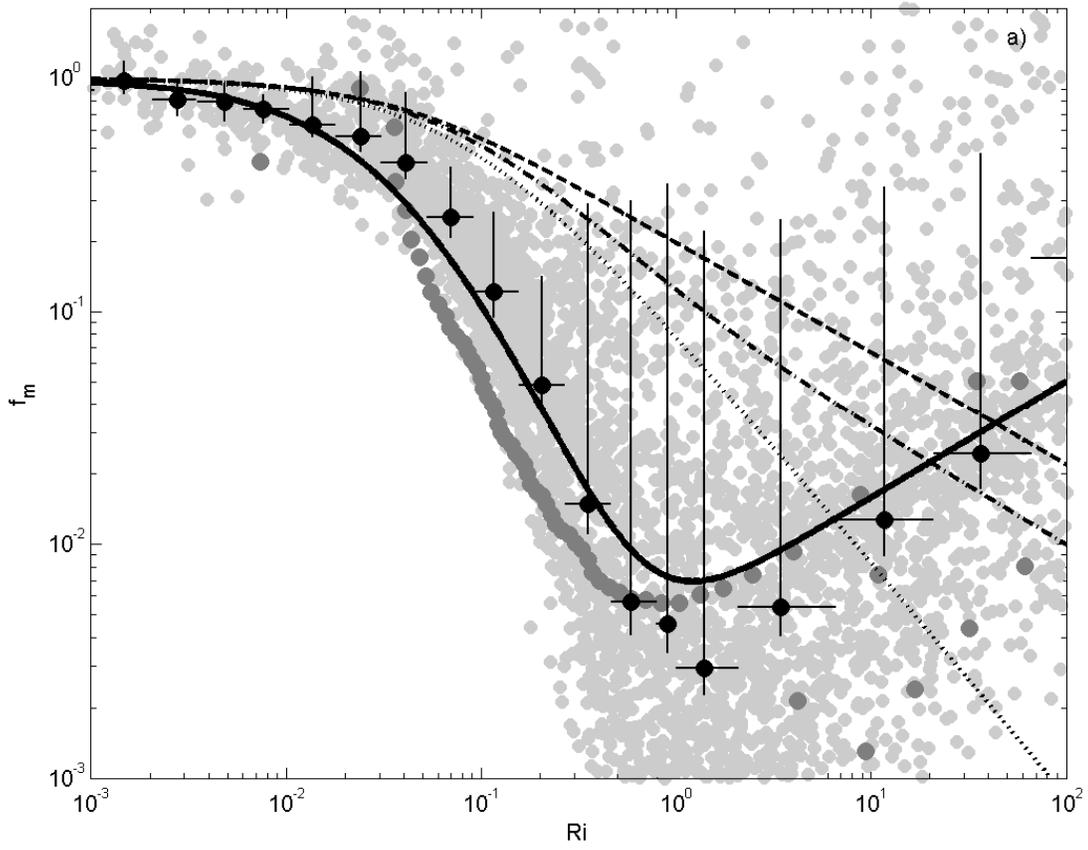

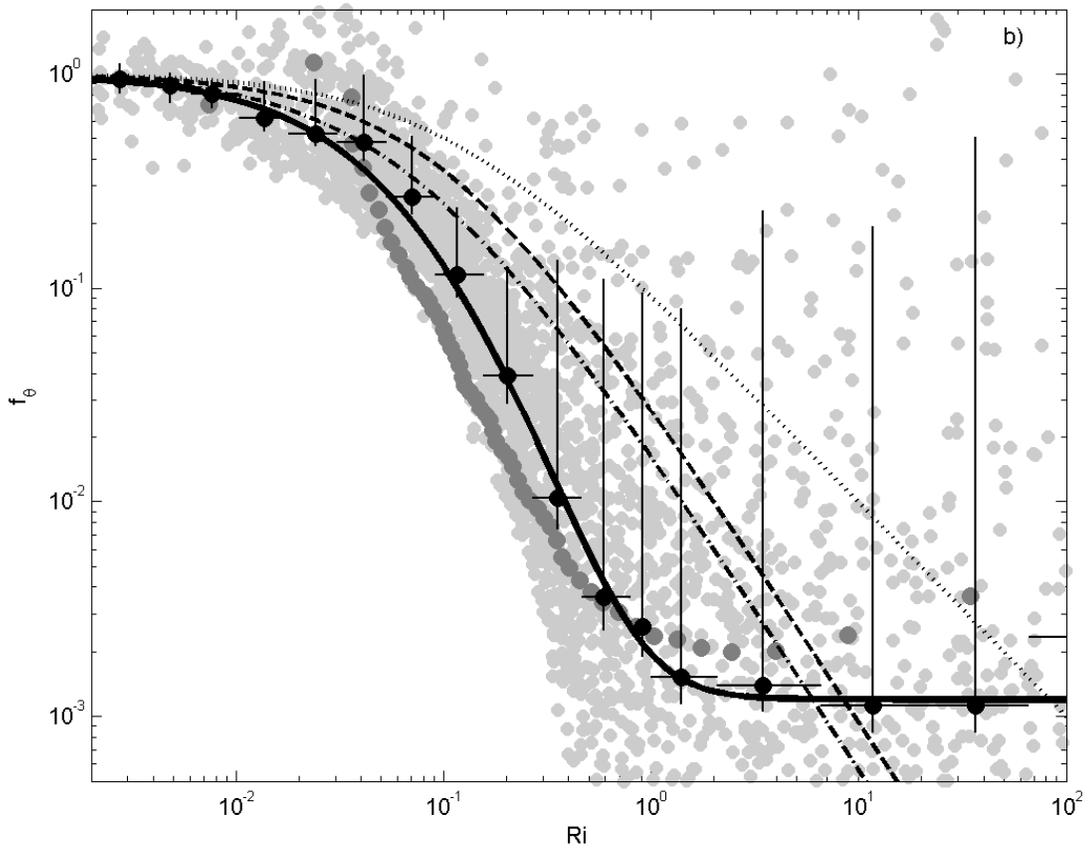

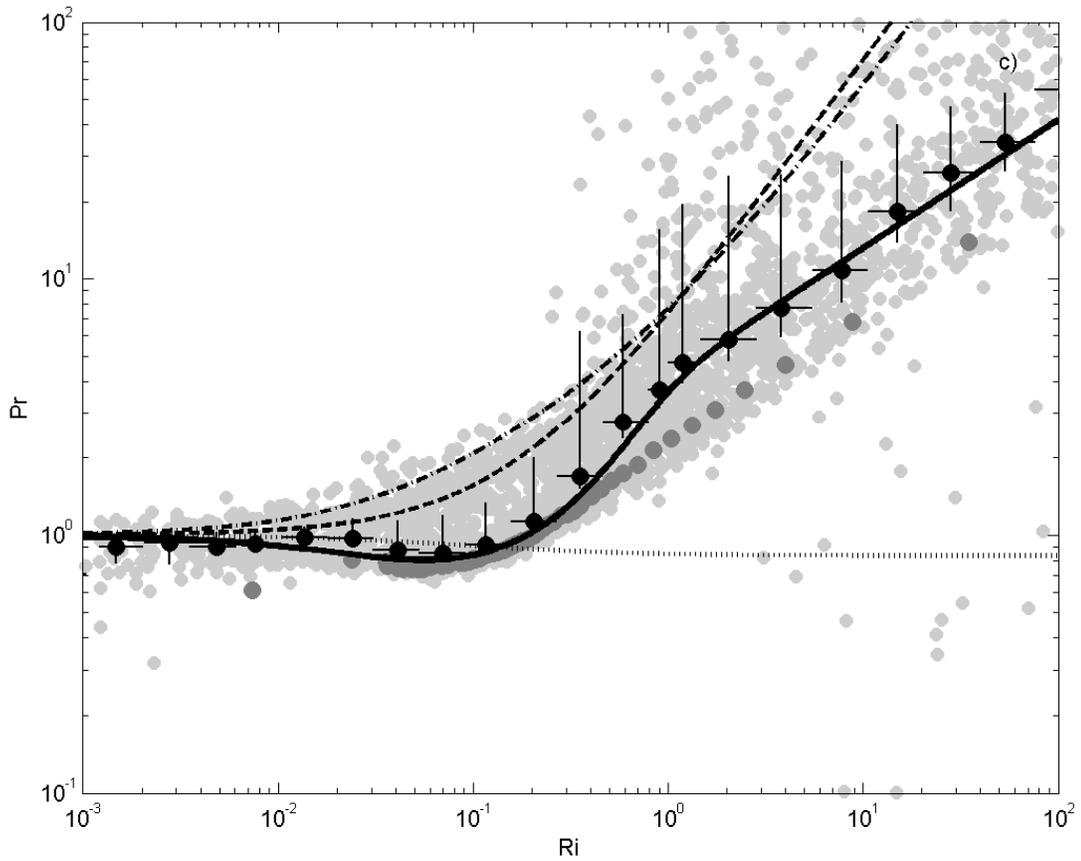

**Figure 4.** Stability correction functions and their ratio (the Prandtl number) as functions of the gradient Richardson number: $f_m(Ri)$ – for momentum (a); $f_h(Ri)$ – for temperature (b); and $\Pr = f_m(Ri)/f_h(Ri)$ – the Prandtl number (c). Symbols are: light dots – DATABASE64 data; dark dots – LES simulations for the GABLS case (Beare et al., 2006); black dots with error bars – bin-averaged values of equally weighted DATABASE64 data, the horizontal bars denote the bin width, the vertical bars denote one standard deviation within each bin for the data found above and below of the bin-average value. Curves are: solid curve – the best fit to the bin-averaged DATABASE64 data; dashed curve – the approximation used in the ECMWF, the ARPEGE (MeteoFrance) and some other models (Louis, 1982; C06); dash-dotted curve – an improved MeteoFrance approximation discussed in C06; dotted curve – the approximation used in the MetOffice model.

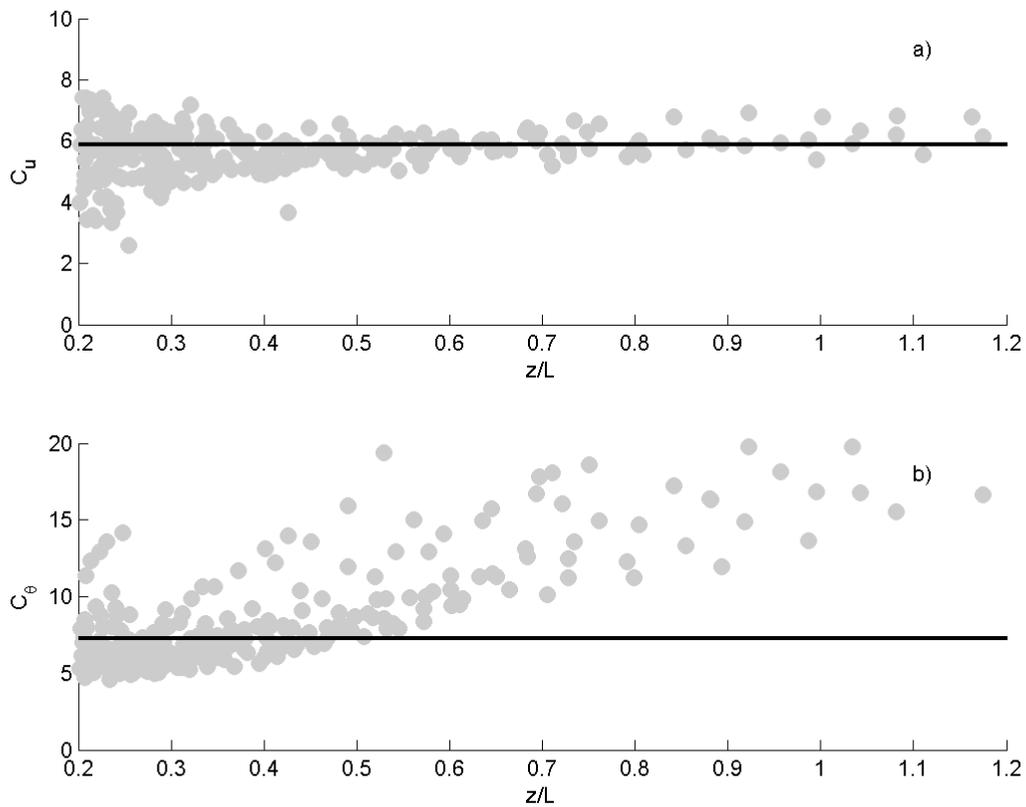

**Figure 5.** Constants in flux-profile relationships after Eq. (3b): $C_u$ – for momentum (a) and $C_\theta$ – for the temperature (b) non-dimensional gradients as functions of the height normalized by the surface Monin-Obukhov length scale. Light dots denote DATABASE64 data for the lower 1/3 of the PBL with exception of the surface layer (1-2 computational levels in the LES). The bold lines are the mean values of the constants $C_u$ = 5.9 and $C_\theta$ = 7.3. It is clearly seen that the linear approximation for the non-dimensional temperature gradient is not suitable.

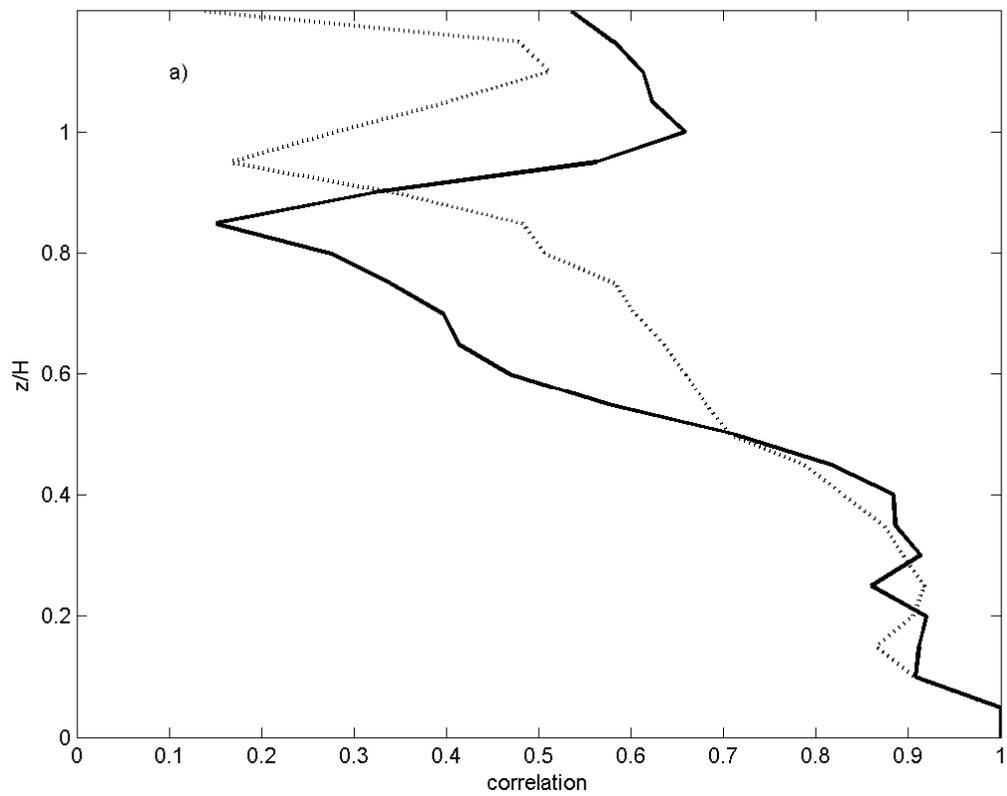

b)
z/H

correlation

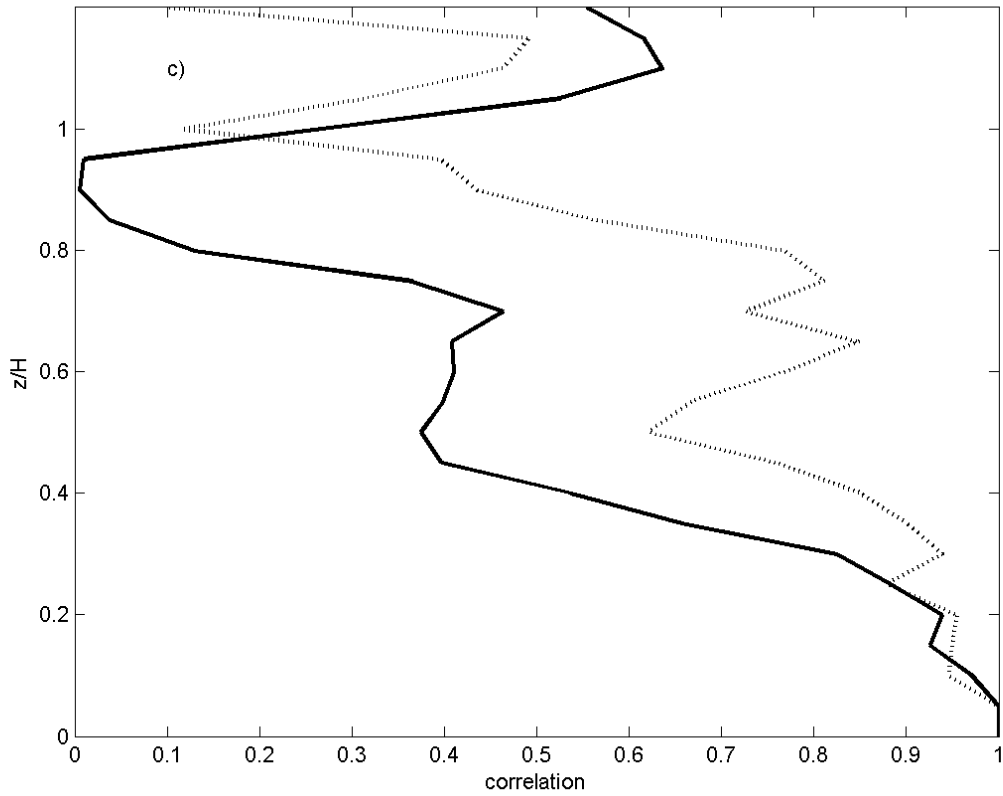

**Figure 6.** Vertical profiles of the correlations between the turbulent fluxes in the optimized FOC and the directly computed LES fluxes from DATABASE64. Solid curve is for the momentum and the dotted curve – for the temperature flux correlations: (a) – averaged over all cases; (b) – averaged over conventionally neutral cases; (c) – averaged over nocturnal cases. High correlations near the surface, $z/H < 0.05$, are artificial product of the numerical analysis.

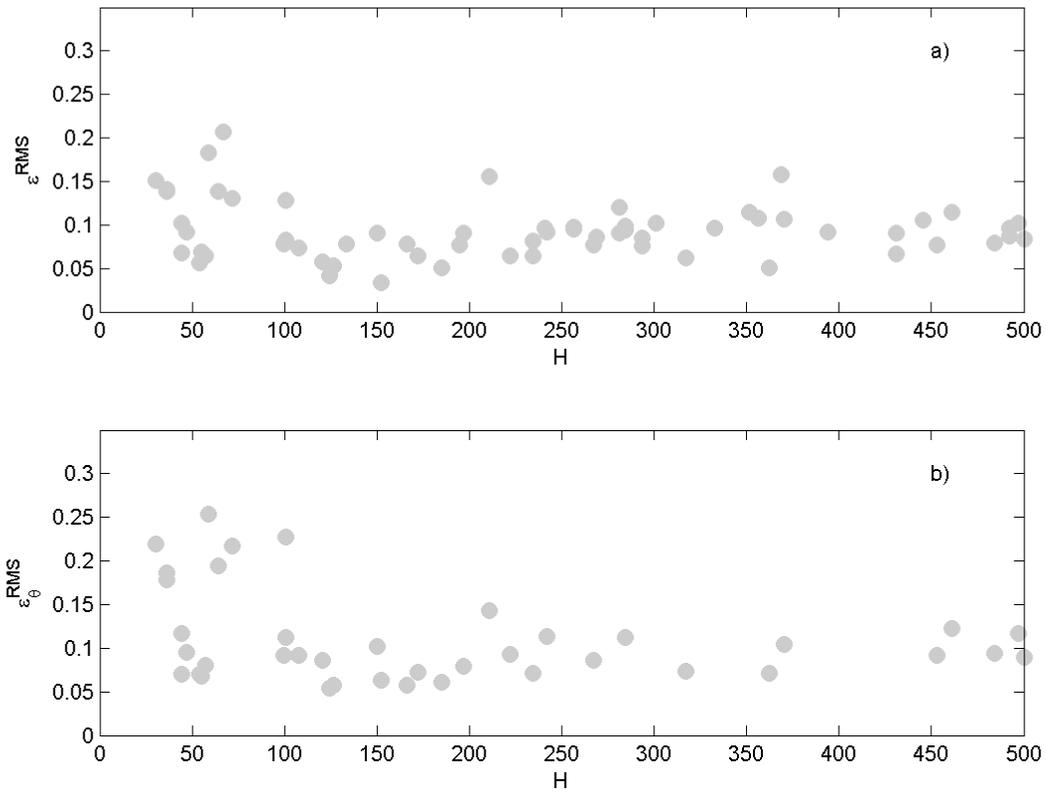

**Figure 7.** Relative root-mean square error over the boundary layer after Eq. (9) for the momentum (a) and temperature (b) fluxes computed for the optimized FOC from DATABASE64 wind and temperature profiles.

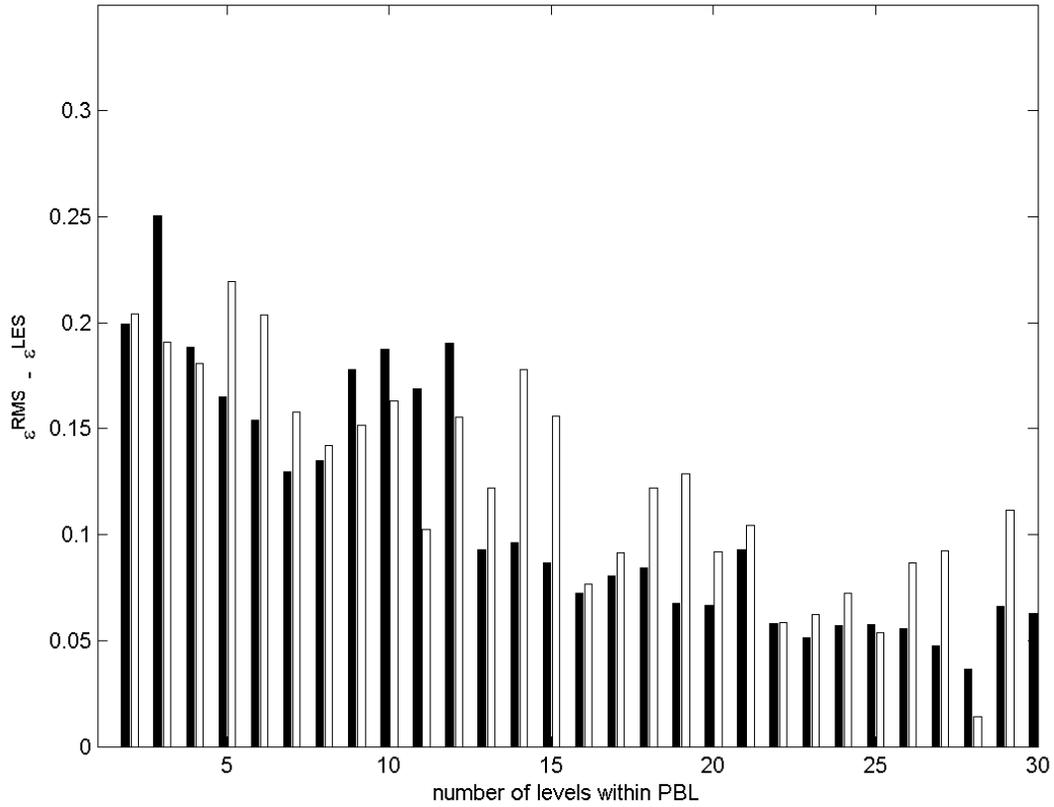

**Figure 8.** The FOC quality dependence on model resolution within the PBL. The quality computed after Eq. (9) with subtraction of the minimum error for every LES run. Dark bars are for the momentum flux and light bars – for the temperature flux computed by the optimized FOC from the DATABASE64 wind and temperature profiles.

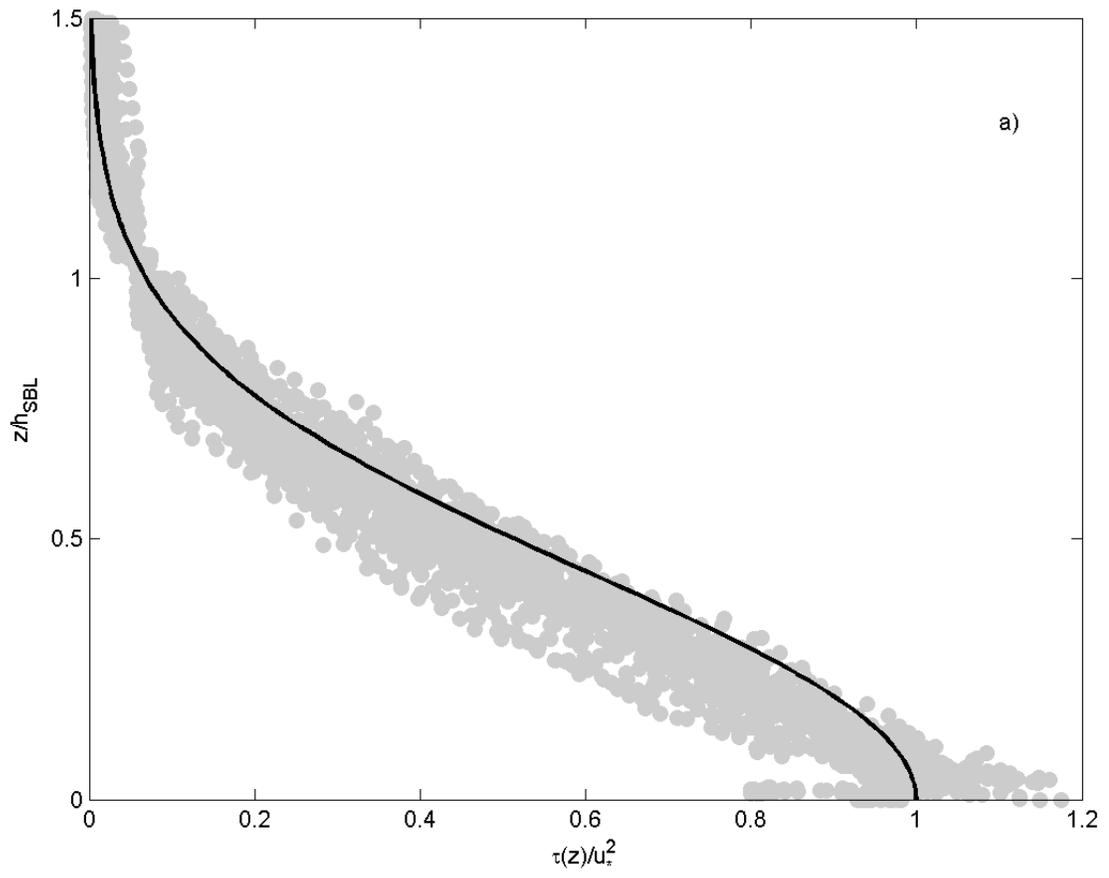

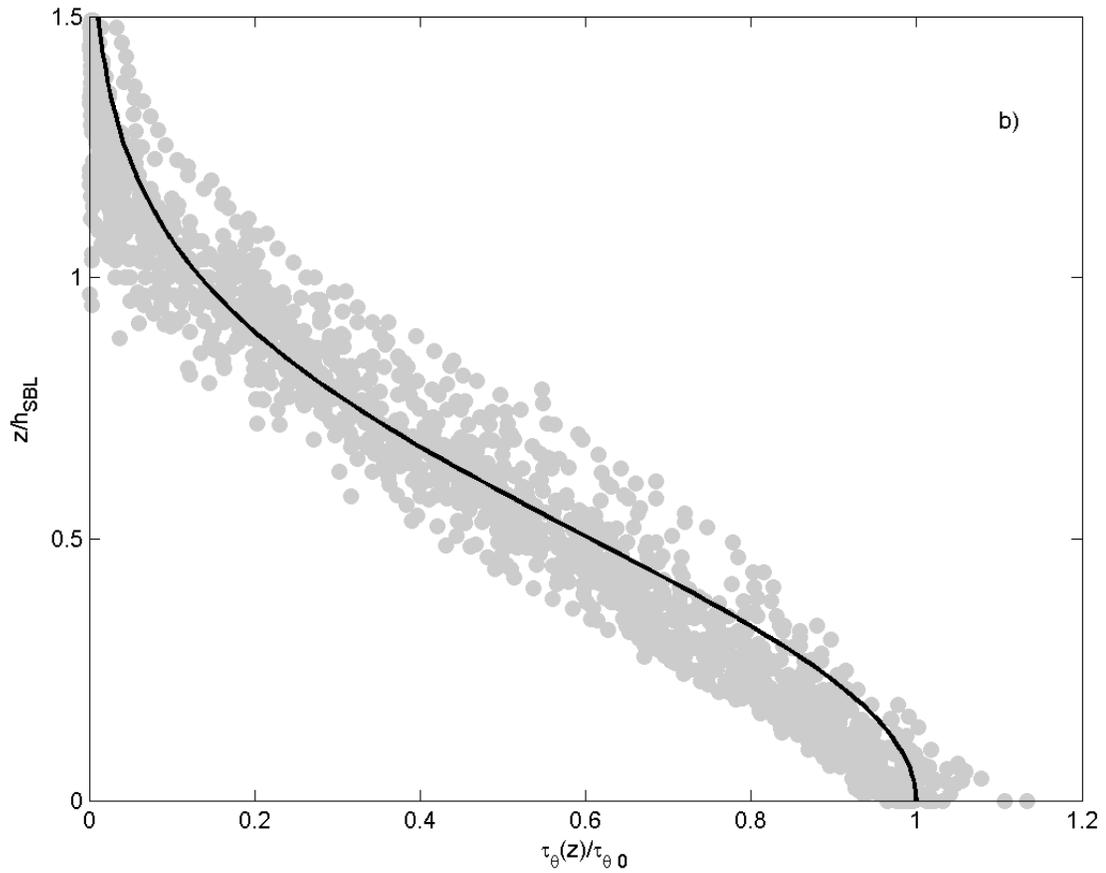

**Figure 9.** Non-dimensional turbulent fluxes for momentum (a) and temperature (b) taken from DATABASE64 (light dots) and their analytical approximations (bold curves) with Zilitinkevich and Esau (2005) exponential universal functions after Eq. (11).

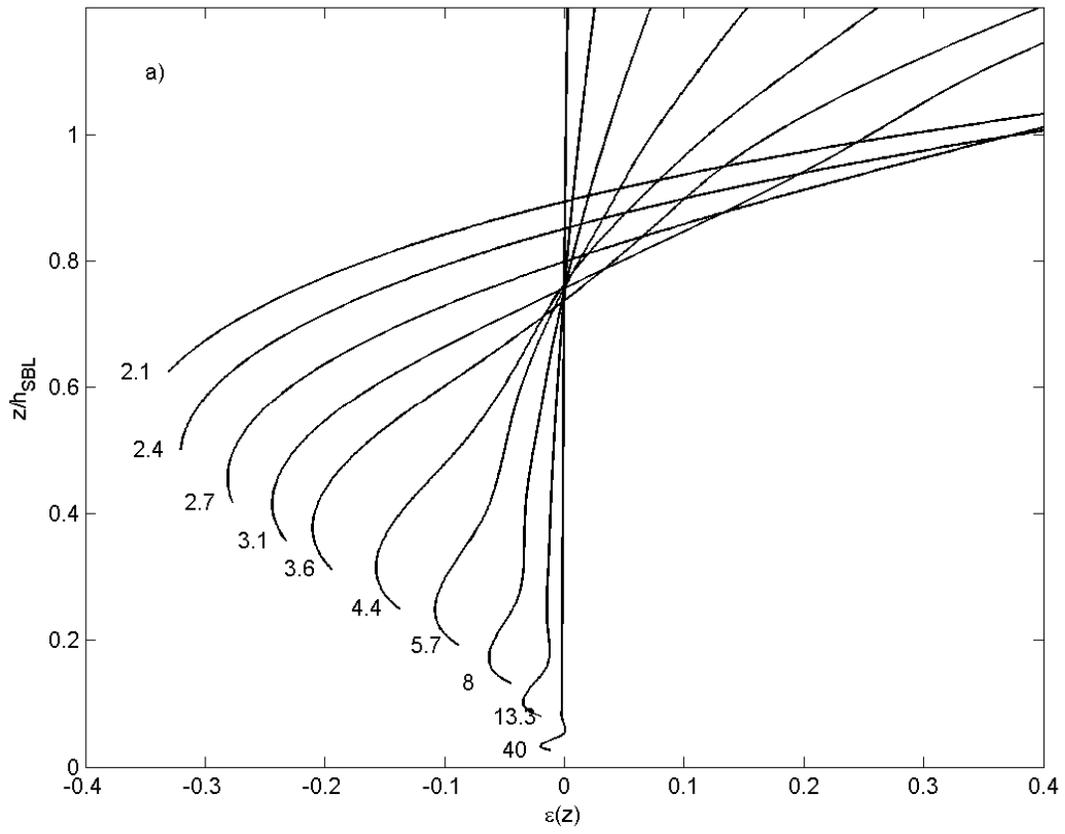

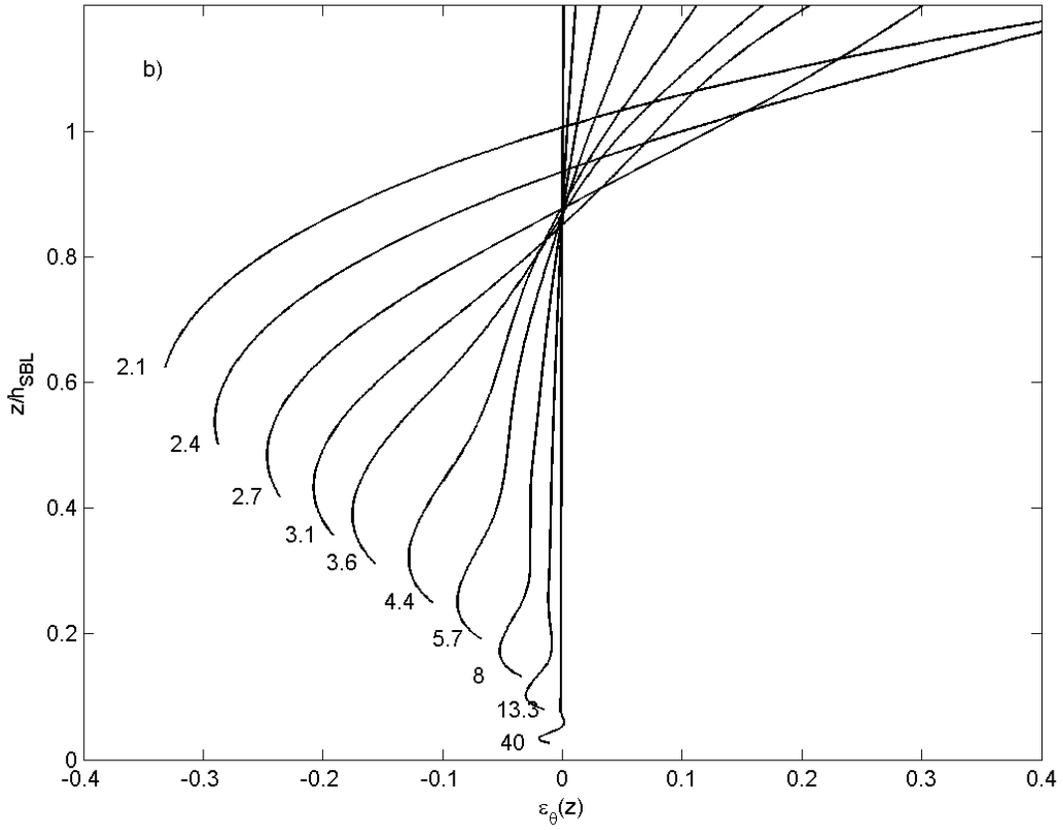

**Figure 10.** Profiles of relative errors in momentum (a) and temperature tendencies (b). The errors are computed by Eq. (12) for different equidistant meshes with $N_z = H/\delta$ indicated at the bottom of the corresponding curve.

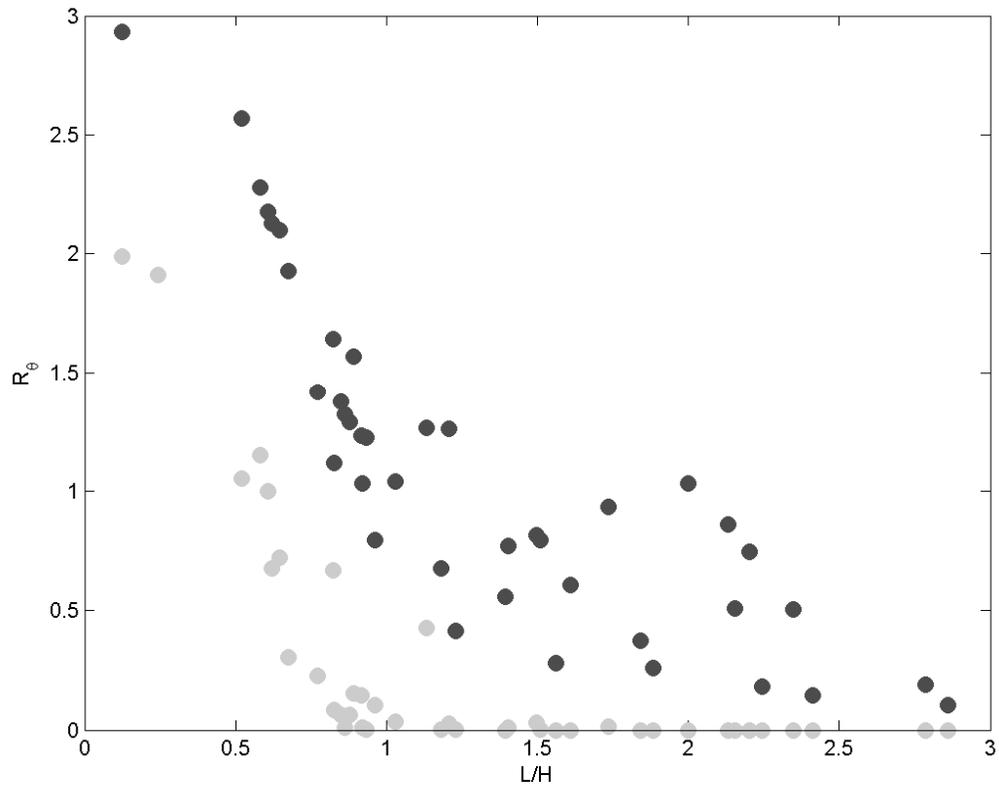

**Figure 11.** Ratio between the total and the down-gradient fluxes computed from DATABASE64 after Eq. (13). The ratio is plotted versus non-dimensional ratio between the surface Monin-Obukhov length scale and the PBL thickness. Symbols are: light dots – the averaged values for the lower half of the PBL; dark dots – the averaged values for the upper half of the PBL.

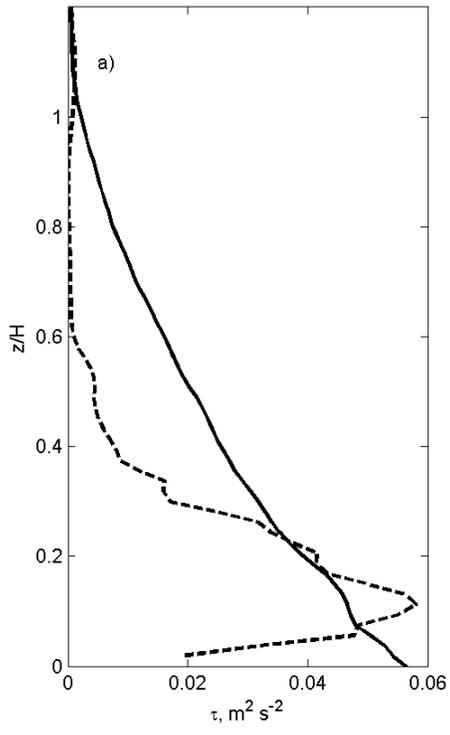
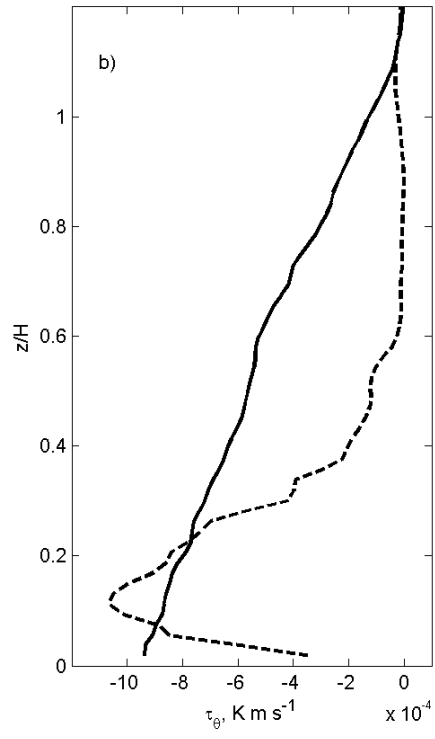
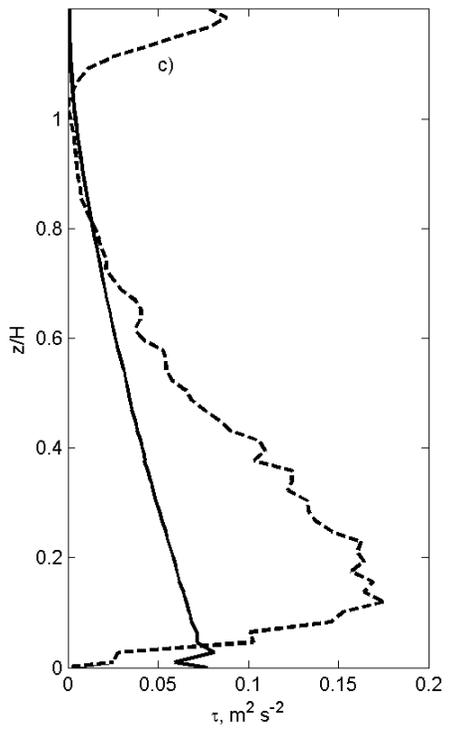
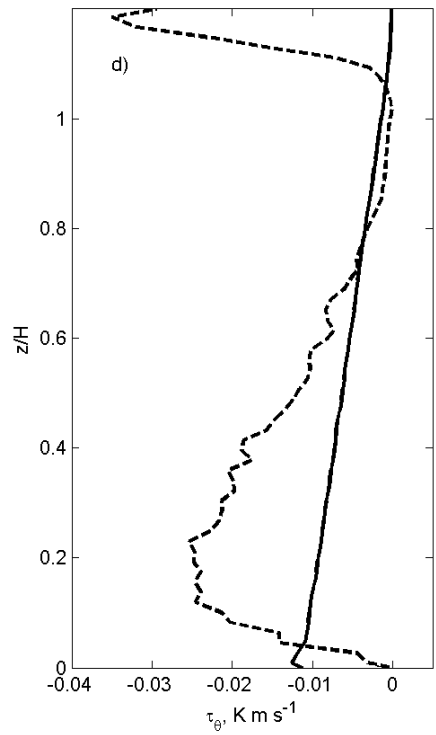

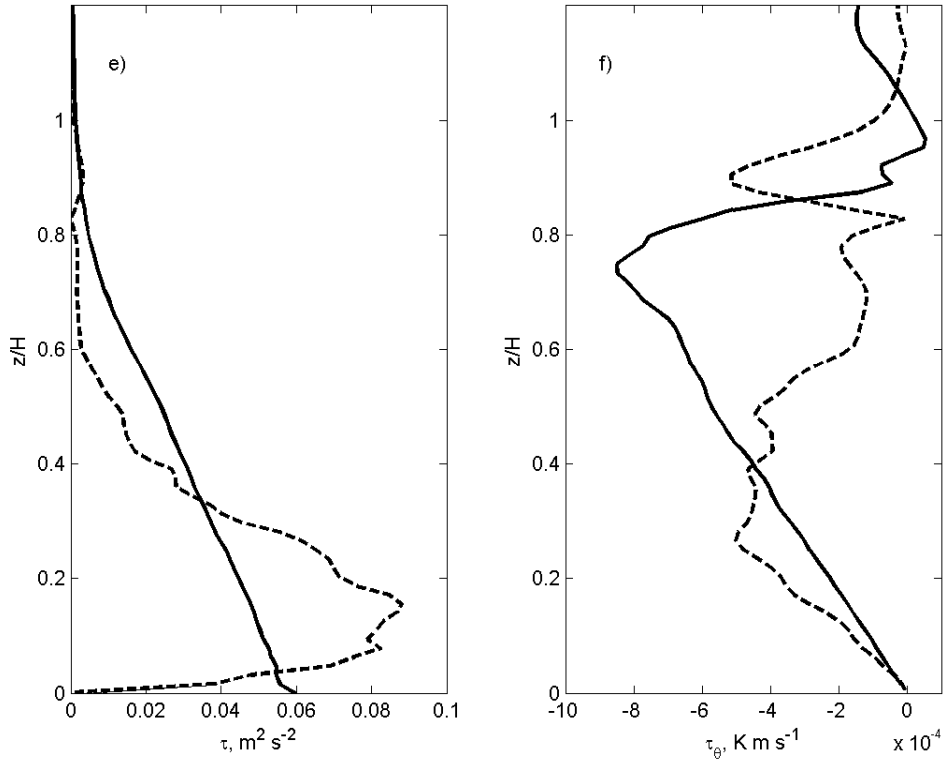

**Figure 12.** Vertical profiles of the LES (solid curves) and the FOC (dashed curves) fluxes for momentum (a, c, e) and temperature (b, d, f) for three distinct cases: (a, b) – the nocturnal PBL with zonal geostrophic wind, $U_g = 5$ m s$^{-1}$, the surface temperature flux $\tau_{\theta 0} = 0.001$ K m s$^{-1}$, at latitude 45 degrees North and surface roughness 0.1 m; (c, d) – the long-lived stably stratified PBL in the GABLS case, with zonal geostrophic wind, $U_g = 8$ m s$^{-1}$, the variable surface temperature flux with the mean value of $\tau_{\theta 0} \sim 0.07$ K m s$^{-1}$, at latitude 73 degrees North and surface roughness 0.1 m; (e, f) – the conventionally neutral PBL with zonal geostrophic wind, $U_g = 5$ m s$^{-1}$, the surface temperature flux $\tau_{\theta 0} = 0$ K m s$^{-1}$, at latitude 45 degrees North and surface roughness 0.1 m. The FOC fluxes in $z/H < 0.05$ are numerical artefact as no surface flux parameterization has been applied.